# Quantifying impacts of the drought 2018 on European ecosystems in comparison to 2003

**Allan Buras[1], Anja Rammig[1], Christian S. Zang[1]**

[1*]Technical University of Munich, TUM School of Life Sciences Weihenstephan, Professorship for Land Surface-Atmosphere Interactions, Hans-Carl-von-Carlowitz Platz 2, 85354 Freising, Germany.

**\*Correspondence:**
Allan Buras

allan@buras.eu

**Keywords: Remote sensing, Compound events, Drought legacy, Ecosystem productivity, Climatic water balance, Atmospheric circulation patterns.**

**Updates in pre-print version 2 (submitted on July 15$^{th}$ 2019):**

We now apply an individual detrending (instead of a uniform detrending) to MODIS VI time-series (see section 2.1.2 and supplementary Fig. S1) which results in a stronger drought response of VI in 2018 compared to 2003. While the main conclusions have not changed, specific values (such as the area featuring lowest quantiles in 2018 but also explained variances of models) have changed in comparison to version 1.

**Abstract**

In recent decades, an increasing persistence of atmospheric circulation patterns has been observed. In the course of the associated long-lasting anticyclonic summer circulations, heat waves and drought spells often coincide, leading to so-called hotter droughts. Previous hotter droughts caused a decrease in agricultural yields and increase in tree mortality, and thus, had a remarkable effect on carbon budgets and negative economic impacts. Consequently, a quantification of ecosystem responses to hotter droughts and a better understanding of the underlying mechanisms is crucial. In this context, the European hotter drought of the year 2018 may be considered a key event. As a first step towards the quantification of its causes and consequences, we here assess anomalies of atmospheric circulation patterns, temperature loads, and climatic water balance as potential drivers of ecosystem responses as quantified by remote sensing using the MODIS vegetation indices NDVI and EVI. To place the drought of 2018 within a climatological context, we compare its climatic features and ecosystem response with the extreme hot drought of 2003. Our results indicated 2018 to be characterized by a climatic dipole, featuring extremely hot and dry weather conditions north of the Alps but comparably cool and moist conditions across large parts of the Mediterranean. Analyzing ecosystem response of five dominant land-cover classes, we found significant positive effects of April-July climatic water balance on ecosystem productivity. Negative drought impacts appeared to affect a larger area and to be significantly stronger in 2018 compared to 2003. Moreover, we found a significantly higher sensitivity of pastures and arable land to climatic water balance compared to forests in both years. The stronger coupling and higher sensitivity of ecosystem response in 2018 we explain by the prevailing climatic dipole: while the generally water-limited ecosystems of the Mediterranean experienced above-average climatic water balance, the less drought-adapted ecosystems of Central and Northern Europe experienced a record hot drought. In conclusion, this study quantifies the drought of 2018 as a yet unprecedented event and provides valuable insights into the heterogeneous responses of European ecosystems to hotter drought.

## 1   Introduction

More frequent and longer-lasting heat waves are expected to occur with global warming (IPCC, 2014). If such heat waves coincide with low precipitation sums, so-called 'global-change type droughts' or 'hotter droughts' emerge (Allen et al., 2015; Breshears et al., 2005). In the course of hotter droughts, positive feedback loops related to a non-linearly amplified soil-water depletion through evapotranspiration (Seneviratne et al., 2010) further aggravate surface-temperature anomalies because of reduced latent cooling (Fischer et al., 2007). To emphasize this interdependence of heat and drought and improve projections of potential high-impact events, hotter droughts were recently classified as compound events (Zscheischler et al., 2018). Accounting for the interdependence of climatic drivers for drought, climate model projections generally indicate an increase in the likelihood of a hotter drought during the 21$^{st}$ century (Zscheischler and Seneviratne, 2017). Given the associated climatic properties, hotter droughts are more likely to occur under abnormally stable anticyclonic atmospheric circulation patterns which were recently shown to be connected with a hemisphere-wide wavenumber 7 circulation pattern (Kornhuber et al., 2019). Abnormally stable anticyclonic atmospheric circulation patterns and associated wavenumber 7 circulation patterns have expressed an increasing frequency over the past decades (Horton et al., 2015; Kornhuber et al., 2019).

Hotter droughts feature a wide range of negative impacts on managed and natural ecosystems, e.g. reduced productivity, which has been indicated by lower vegetation greenness using remote sensing data (Allen et al., 2015; Choat et al., 2018; Ciais et al., 2005; Orth et al., 2016; Xu et al., 2011). As a consequence, agricultural yields decline remarkably during hotter droughts while drought-induced tree mortality increases, with both effects leading to significant economic losses (Allen et al., 2010; Buras et al., 2018; Cailleret et al., 2017; Choat et al., 2018; Ciais et al., 2005; Matusick et al., 2018). Moreover, since gross primary productivity (GPP) decreases during hotter droughts, the resulting lower net carbon uptake may change ecosystems from carbon sinks into carbon sources (Ciais et al., 2005; Xu et al., 2011). However, the response to drought may vary among different land-cover types, particularly between grasslands and forests (Teuling et al., 2010; Wolf et al., 2013).

On the continental scale, the European heat wave of 2003 is to date considered as the most extreme compound event in Europe with various impacts on human health (increased mortality particularly in France), economy (decreased crop yield in agriculture and forestry), and ecosystems (reduced productivity, forest die-back, and an increased frequency of forest fires; Fink et al., 2004; García-Herrera et al., 2010). According to Ciais et al. (2005), GPP of European ecosystems was reduced by 30 percent in summer 2003 – a yet unprecedented reduction in Europe's primary productivity which resulted in an estimated net carbon release of 0.5 PG C yr-1. Given the wide-ranging impacts, potential climate change feedback loops, and the increasing frequencies of circulation patterns initiating compound events it is pivotal to better understand and thus more precisely predict the response of managed and natural ecosystems to hotter droughts (Horton et al., 2015; Pfleiderer and Coumou, 2018; Sippel et al., 2017; Zscheischler and Seneviratne, 2017).

In the context of an increased persistence of circulation patterns, the European drought of 2018 is of particular interest. In April 2018, a high-pressure system established over Central Europe and persisted almost continuously until mid of October, thereby causing a long-lasting drought spell and record temperatures in central and northern Europe. Despite preliminary reports in public news and the world-wide-web (see list of public news references), the direct impacts resulting from the 2018 drought are still unexplored. Consequently, we here quantify the impacts of the extreme drought of 2018 on European ecosystems in comparison to the extreme drought in the year 2003. Thereby, we 1) provide an estimate of European ecosystems immediate response to the drought 2018 in relation to 2003, 2) identify hotspots of extreme drought and associated ecosystem response, and 3) aim at an improved mechanistic understanding of the processes driving ecosystem responses to extreme drought events.

## 2  Material and Methods

### 2.1 Data sources and preparation

### 2.1.1 Climate data

To visualize the general circulation patterns in 2003 and 2018 we downloaded gridded reanalysis data representing 500 hPa geopotential height from the NCEP/NCAR Reanalysis project provided by the NOAA climate prediction center (Kalnay et al., 1996) available at the Earth System Research Laboratory (ESRL, https://www.esrl.noaa.gov/). The downloaded data cover the period 1981-2018 at a daily temporal resolution and a spatial resolution of 2.5°. As a representation of high-pressure persistence, we computed the mean geopotential height for each grid cell and year for the period from 1$^{st}$ of April until 31$^{st}$ of July.

From ESRL, we furthermore downloaded reanalyzed (NCEP/NCAR), daily gridded mean minimum and maximum temperature ($T_{min}$, $T_{max}$) and precipitation (P) sums at 0.5° spatial resolution covering the period 1981-2018 (Kalnay et al., 1996). These variables were used to compute potential evapotranspiration (PET, as defined by Hargreaves, 1994) and the climatic water balance (CWB = P-PET, Thornthwaite, 1948). As for geopotential height, we for each grid cell and year integrated $T_{max}$ and CWB for the period from 1$^{st}$ of April until 31$^{st}$ of July as measures of heat load and water balance.

Processed climate data were spatially truncated to match the region considered for the MODIS satellite images (see next section) resulting in 2312 climate grid cells representing an area of roughly 5.9 million km² and covering 38 years. To allow for combination with MODIS data throughout the analyses, processed climate data were re-projected to MODIS native projection using zonal means while retaining a spatial resolution of 0.5°.

To locally quantify climatic conditions, we furthermore extracted monthly temperature means and precipitation sums for six climate stations from the European Climate Assessment and Data project (Tank et al., 2002) available at the corresponding project webpage ([www.ecad.eu](www.ecad.eu)). We chose Oslo and Stockholm, Amsterdam and Berlin, as well as Madrid and Sevilla to represent weather conditions in Northern, Central, and Southern Europe, respectively. Monthly temperature means and precipitation sums of 2003 and 2018 were visually compared to average values representative of the climate normal period 1961-1990.

### 2.1.2 *MODIS* vegetation indices

Using the Application for Extracting and Exploring Analysis Ready Samples (AppEEARS; [https://lpdaacsvc.cr.usgs.gov/appeears](https://lpdaacsvc.cr.usgs.gov/appeears)) we downloaded two MODIS vegetation indices (VI, i.e. the Normalized Difference Vegetation Index NDVI and the Enhanced Vegetation Index EVI) and the corresponding pixel reliability layers at 231 m spatial resolution and 16 day temporal resolution in their native projection. The downloaded data cover the area between 10° E and 30° W longitude, 36.5° N and 71.5° N latitude that is represented by the CORINE land cover information of 2012 (see section 2.1.3) and span the period from February 2000 until end of 2018.

Based on the pixel reliability information, we only retained records with good or marginal quality for subsequent analyses. Consequently, for most of the grid cells the VI time series contained missing values due to temporary clouds or snow cover. If the number of missing values was larger than the number of VI records, we considered the representing records as insufficient for our analyses and consequently removed the corresponding pixel from the analysis. However, since high-elevation as well as high-latitude pixels had many missing values in winter and spring because of clouds and snow cover and we only were interested in VI during peak season, we only considered the period from beginning of March (DOY 64) to end of October (DOY 304) for the definition of valid pixels. Following these selection criteria, we retained 95,523,236 pixels for the final analyses, representative of an area of 5,970,202 km².

Prior to the analyses, VI time series of the retained pixels were further processed. We linearly interpolated the missing values of the corresponding VI time series for each pixel using the previous and succeeding records (Misra et al., 2016, 2018). Subsequently, we removed negative outlier values from each VI time series by computing standardized residuals to a Gaussian-filtered (filter size of 80 days, i.e. 5 MODIS time steps), smoothed time-series. Residuals exceeding two negative standard deviations were replaced by the equivalent value of the smoothed time series (see also Misra et al., 2018, 2016). We smoothed the interpolated, outlier-corrected time series by reapplying the Gaussian filter. This procedure was necessary to efficiently handle the remaining high-frequency variability in the seasonal VI-cycle (Misra et al., 2016, 2018).

Finally, VI time series were detrended for each pixel individually by determining the linear trend of VI for each pixel and subtracting the pixel-specific trend from the corresponding pixel time series. This detrending was necessary to compensate for the observed trends in vegetation indices (Bastos et al., 2017) which were also apparent in the downloaded data (Fig. S1). A comparison between non-detrended and detrended data revealed similar spatial patterns with respect to between-pixel variability, however with amplified differences between 2003 and 2018 in the raw, non-detrended data. That is, for the raw data, the observed trend caused lower peak-season VI values in 2003 compared to 2018, thereby introducing an offset between these two drought events. Concluding, the detrending was able to efficiently handle the VI-trend over the MODIS-era, while spatial patterns were generally retained.

Both NDVI and EVI are considered as proxy for photosynthetic carbon fixation, and thus allow for assessing possible changes in productivity in dependence of environmental conditions (Huete et al., 2006; Myneni et al., 1995; Xu et al., 2011). Since NDVI previously has been used in the context of drought monitoring (Anyamba and Tucker, 2012), we focus on results derived from NDVI which are generally supported by EVI. To allow for a direct comparison of NDVI and EVI results, the latter are depicted in the supplementary.

### 2.1.3 *CORINE* land cover information

To get an impression on the drought-impact on key European ecosystem components, analyses were stratified using the Coordinated Information on the European Environment land cover map (CORINE, https://land.copernicus.eu/pan-european/corine-land-cover/clc-2012) at 250 m spatial resolution. The land cover map was re-projected (as were the gridded climate data) to MODIS native projection using the nearest neighbor method, thereby retaining the original land-cover classes. Given their dominance in Europe and their importance for land-use, we constrained this stratification to pastures, arable land, as well as coniferous, mixed, and broadleaved forests.

### 2.2 Statistical analyses

To quantify weather conditions for the years 2003 and 2018 in relation to average conditions, standardized anomalies of 500 hPa geopotential height, heat load and CWB were calculated. Before doing so, we tested the underlying assumption of normal distribution by computing Shapiro test for each grid cell and climate parameter respectively (Fig. S2). The number of significant tests ($p < 0.001$) indicating non-normal distribution was in the order of expected false positives (0.0-0.3 percent vs. 0.1 percent type I error probability). Thus, we considered the assumption of normality to be generally fulfilled. To derive anomalies, we first computed the mean and standard deviation for all variables for the full period (1981-2018). Subsequently, we for 2003 and 2018 determined the difference of the respective metric to its corresponding mean in units of standard deviations which in the following are called standardized anomalies (this procedure is also known as z-transformation). Thus, for integrated geopotential height, heat load, and CWB we obtained each one standardized anomaly per grid cell for 2003 and 2018. The resulting standardized anomalies were mapped and statistically evaluated using histograms. Histograms were used to depict the absolute area of differently affected regions in 2003 and 2018 which were compared among each other as well as to a normal distribution. This was done in order to visualize the severity of these two drought events in comparison to each other as well as in comparison to normal conditions.

To quantify the response of European ecosystems to the two drought events, we focused on end-of-July (DOY 209) VI values. The selection of this particular date represents a compromise between proximity to peak-season (end of June, before maximum temperatures had been reached) and the occurrence of heat-waves (end of July to mid of August). Since VI features a bounded distribution (values between -1 and +1), we could not apply a standardization approach as for the climate variables. Therefore, we for each VI time series computed its end-of-July quantiles over the 19 years similar to Orth et al. (2016). The corresponding quantiles were mapped for 2003 and 2018. Areas representing the 19 different quantiles were extracted and compared between 2003 and 2018 in a histogram.

Since we were aiming at a better understanding of particular ecosystems' response to drought severity, we subsequently pooled VI quantiles according to three classes of CWB anomalies (abnormal water deficit: CWB < -2, average water supply: -2 < CWB < 2, abnormal water surplus: CWB >2) and five CORINE land-cover classes (arable land, pastures, coniferous forest, mixed forest, broadleaved forest). For the resulting 15 combinations, we again compared the areas representing the 19 different quantiles between 2003 and 2018 as done for the total scene. Since the areas of CWB-land cover combinations differed between 2003 and 2018, we moreover computed histograms expressing proportional areas for 2003 and 2018.

Finally, we aimed at developing empirical relationships between CWB anomalies and VI quantiles for the five CORINE land-cover classes mentioned before. For this, we logit transformed VI-quantiles (quantiles ranging from 0 to 1) to obtain an unbounded distribution and subsequently extracted the corresponding mean of transformed VI-quantiles for each CWB grid-cell (thus n = 2312). To assess the effect of different land cover classes, we extracted both the mean EVI-quantiles representing all five land cover classes as well as for each land cover class separately. For the corresponding 2312 CWB-pixels we computed linear regressions between the transformed EVI-quantiles as the dependent variable and CWB anomalies as independent variable separately for 2003 and 2018 and for the six different land cover types (i.e. five separate classes as well as their combination).

For linear regression evaluation, we report adjusted r² and display scatterplots of VI quantiles vs. CWB along with the corresponding regression line. Moreover, regression slopes were compared statistically for each land cover type between 2003 and 2018. For this, each slope estimate was bootstrapped using random subsampling over 1000 iterations and the overlap of 95 % (99 %) confidence intervals was evaluated. That is, in case the confidence intervals of a respective comparison did not overlap, we considered the difference between slopes as (highly) significant. In a similar manner, we compared model slopes among ecosystems (i.e. pastures, arable land, as well as coniferous, mixed, and deciduous forest) separately for 2003 and 2018. Model slopes were grouped according to their overlap of 95 % confidence intervals. Finally, we also computed a global model by applying a linear mixed effects model (lme) to model EVI quantiles using climatic water balance as fixed effect and incorporating crossed random slopes of land cover and year. All analyses were performed in 'R' (R core team, 2019) extended for the packages, 'nlme' (Pinheiro et al., 2018), 'raster' (Robert J. Hijmans, 2017), and 'SPEI' (Beguería and Vicente-Serrano, 2013).

## 3 Results

All considered climate parameters indicated abnormal weather conditions for 2018 (Figs. 1-3). First of all, the integrated 500 hPa geopotential height, an indicator of the persistence of the atmospheric circulation, expressed anomalies in the order of two positive standard deviations for large parts of Central and Northern Europe, mainly covering the Baltic Sea region (Fig. 1 b). In comparison, 2018 differed from 2003 by featuring a dipole of 500 hPa geopotential height anomalies. While in 2003 most of Europe featured strong positive anomalies, the Mediterranean was characterized by low geopotential height anomalies in 2018 (Fig. 1 a vs. Fig. 1 b). Consequently, the observed dipole of 2018 expressed a bimodal distribution of anomalies whereas 2003 featured a skewed distribution towards positive anomalies (Fig. 1c). The bimodal distribution of 2018 compared to 2003 is also reflected in a 0.55 times as large area featuring positive anomalies, i.e. 3.1 million km² in 2018 vs. 5.7 million km² in 2003, but a 3.7 times higher area with negative anomalies, 3.5 million vs. 0.9 million km² (Fig. 1c).

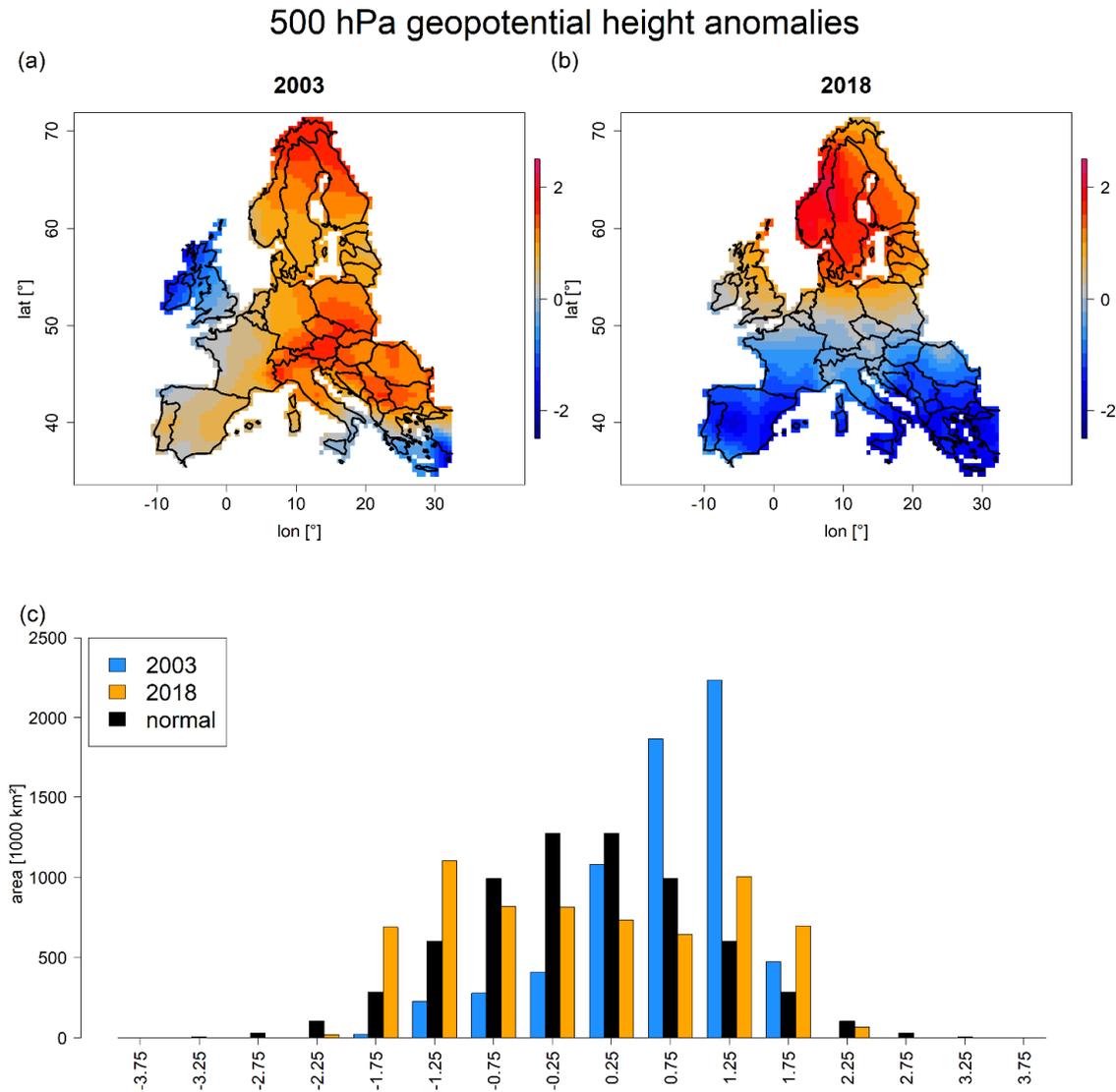

Fig. 1: Maps depicting standardized anomalies of April-July 500 hPa geopotential height for 2003 and 2018 (a, b) as well as corresponding area histograms (in units of 1000 km²) for 2003 (blue), 2018 (orange), compared to a normal distribution (black) (c). Blue colors in (a) and (b) indicate geopotential height lower than average, whereas red colors indicate above average geopotential height in comparison to the 1981-2018 mean.

This picture was underlined by heat-load anomalies, which revealed up to four positive standard deviations over large parts of Central and Northern Europe in 2018 (Fig. 2 b). In contrast, the Mediterranean featured average conditions (i.e. slightly warmer or cooler) as well as strong negative anomalies on the Iberian Peninsula. Although the total area with positive heat-load anomalies was more or less similar in 2003 and 2018 (Fig. 2b vs. 2a), anomalies above two positive standard deviations covered a 7.1 times larger area in 2018, i.e. 3.3 million km² vs. 0.5 million km² in 2003 (Fig. 2c). Most contrasting differences between 2003 and 2018 were observed in Southern Italy (hot in 2003, cool in 2018) as well as Scandinavia and the Baltic Sea region (cool in 2003, hot in 2018). Selected climate station data generally supported the impression of higher (lower) heat loads in 2018 compared to 2003 in Northern (Southern) Europe (Fig. S3a).

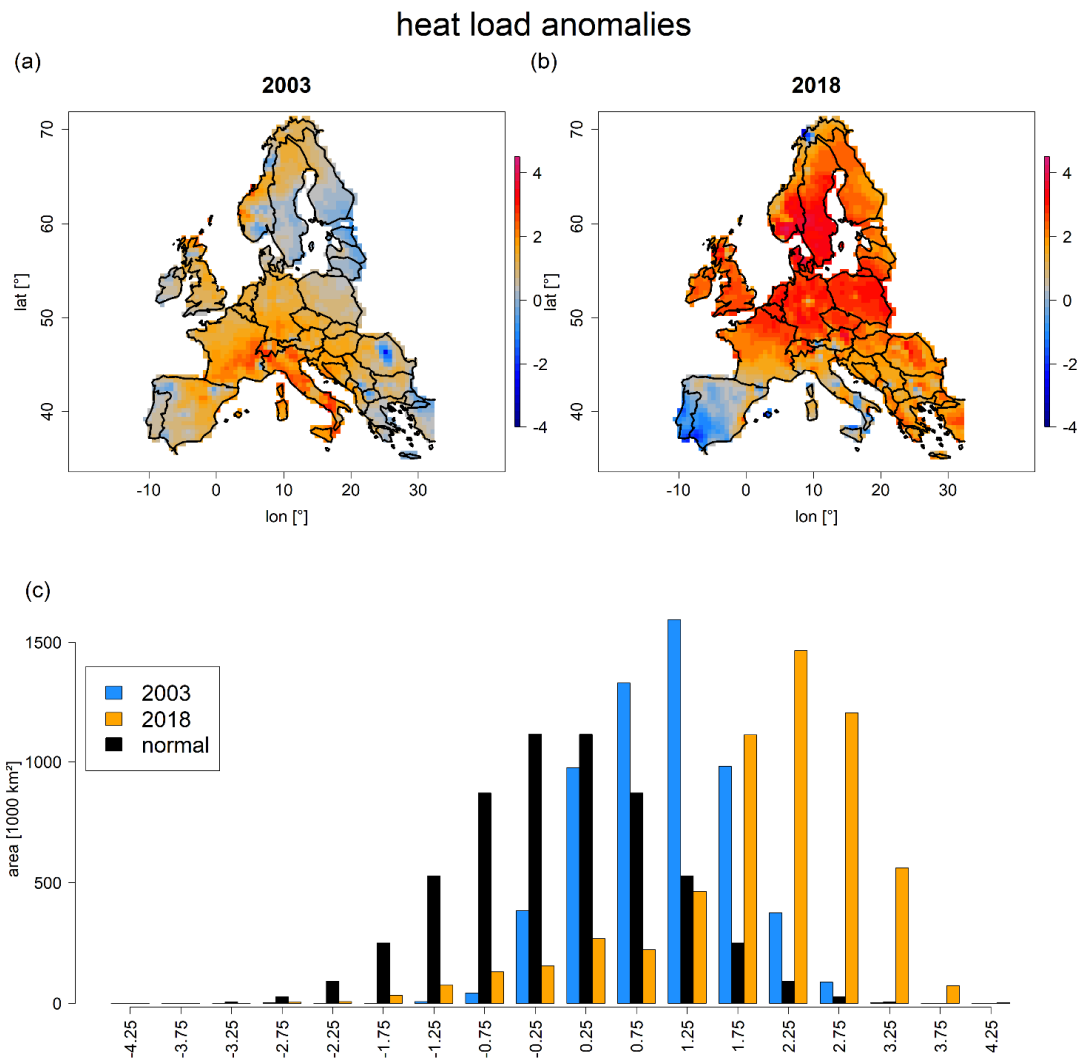

Fig. 2: Maps depicting standardized anomalies of April-July heat load for 2003 and 2018 (a, b) as well as corresponding area histograms (in units of 1000 km²) for 2003 (blue), 2018 (orange), compared to a normal distribution (black) (c). Blue colors in (a) and (b) indicate relatively cool conditions, whereas red colors indicate warmer conditions in comparison to the 1981-2018 mean.

CWB for 2018 revealed patterns largely consistent with heat load (Fig. 3b). Again, Central and Northern Europe featured extreme negative (thus dry) deviations, while the Mediterranean generally expressed positive (thus moist) deviations. In comparison (Fig. 3b vs. 3a), the area with negative (i.e. dry) CWB anomalies was relatively similar in both years, i.e. 4.2 million km² in 2018 vs. 4.6 million km² in 2003 (Fig. 3c). However, when considering only CWB anomalies below two negative standard deviations (i.e. extreme drought), in 2018 an area 5.3 times larger than in 2003 was affected, i.e. 1.7 million km² in 2018 in vs. 0.3 million km² in 2003 (Fig. 3c). Selected climate station data generally supported the impression of higher (lower) water deficit in 2018 compared to 2003 in Northern (Southern) Europe (Fig. S3b).

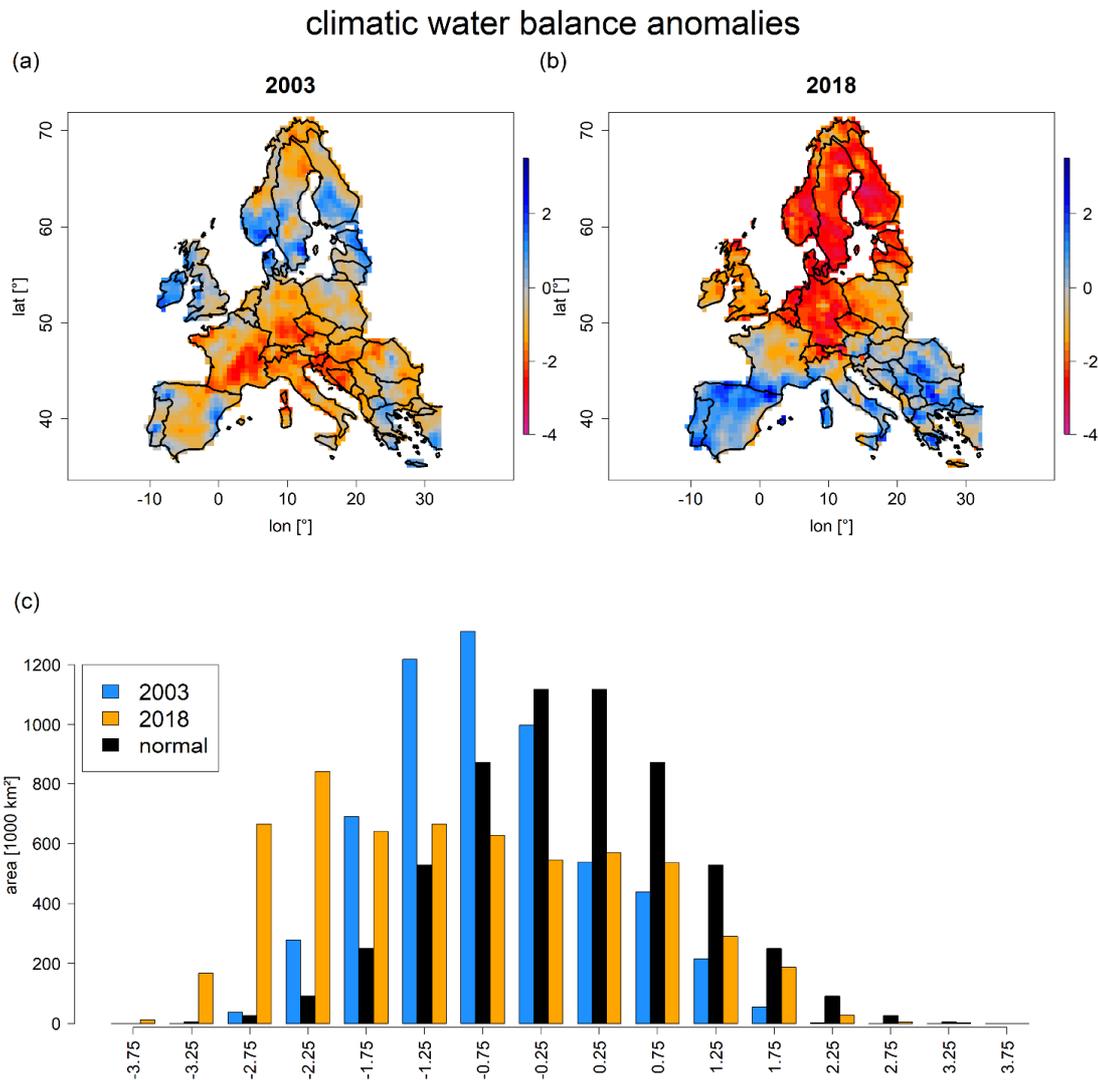

Fig. 3: Maps depicting standardized anomalies of April-July climatic water balance for 2003 and 2018 (a, b) as well as corresponding area histograms (in units of 1000 km²) for 2003 (blue), 2018 (orange), compared to a normal distribution (black) (c). Blue colors in (a) and (b) indicate relatively moist conditions, whereas red colors indicate dryer conditions in comparison to the 1981-2018 mean.

End of July vegetation response indicated clear differences between 2003 and 2018. We found low NDVI quantiles in large parts of Central Europe, Southern Scandinavia, and the Baltic Sea region and high quantiles in the Mediterranean in 2018 (Fig. 4b). In comparison, 2003 featured low NDVI quantiles in Western, Central, and Southeast Europe and high quantiles in Northern Europe (Fig. 4a). The most prominent difference between 2018 and 2003 was the 2 times larger area featuring the lowest quantile, i.e. 822,067 km² in 2018 vs. 408,614 km² in 2003 (Fig. 4c). At the same time, a 2 times larger area featured the highest quantile in 2018, i.e. 439,460 km² vs. 214,922 km² in 2003 (Fig. 4c). According to NDVI quantiles, hotspots of drought-response in 2018 were located in Ireland, United Kingdom, France, Belgium, Luxemburg, the Netherlands, Northern Switzerland, Germany, Denmark, Sweden, Southern Norway, Czech Republic, Poland, Lithuania, Latvia, Estonia, and Finnland.

In regions with water deficit (CWB < - 2; Figs. 5c and S4c) we found a higher frequency of low NDVI quantiles compared to upper quantiles. Similarly, regions with water surplus (CWB > 2; Figs. 5a and S4a) featured higher frequencies for upper quantiles compared to lower quantiles which however was more pronounced in 2018 compared to 2003. Interestingly, normal conditions (Figs. 5b and S4b) featured an inconsistent picture for the different land-cover classes in both years. The most prominent difference was related to absolutely larger areas being affected by water deficit and surplus in 2018 compared to 2003. If considering relative frequencies, histograms of 2018 and 2003 became more similar (Fig. S3).

The impression of CWB affecting NDVI quantiles was underpinned by the linear regressions between the logit-transformed NDVI quantiles and the CWB-anomaly in 2003 and 2018, respectively (Fig. 6a-f). For all land-cover classes a significant and positive effect of climatic water balance on NDVI quantiles was observed, particularly in 2018. Explained variance ($r^2$) and bootstrapped model slopes were consistently higher (though not significantly for pastures) in 2018 compared to 2003 (Fig. 6 g). In addition, $r^2$ and model slopes were in both years highest for pastures, followed by arable land, and the three forest types which did not differ among each other (lower case letters in Fig. 6g). The linear mixed effects model over all land-cover classes and the two years confirmed the significant fixed effect of climatic water balance on logit-transformed NDVI quantiles (marginal $r^2 = 0.47$). Incorporation of random slopes related to land cover and year increased explained variance by 9 percent (conditional $r^2 = 0.51$), confirming the varying effect of the two drought events as well as the differing impact on different land-cover classes. All presented results based on NDVI are supported by complementary analyses using EVI (supplementary Figs. S5-S8).

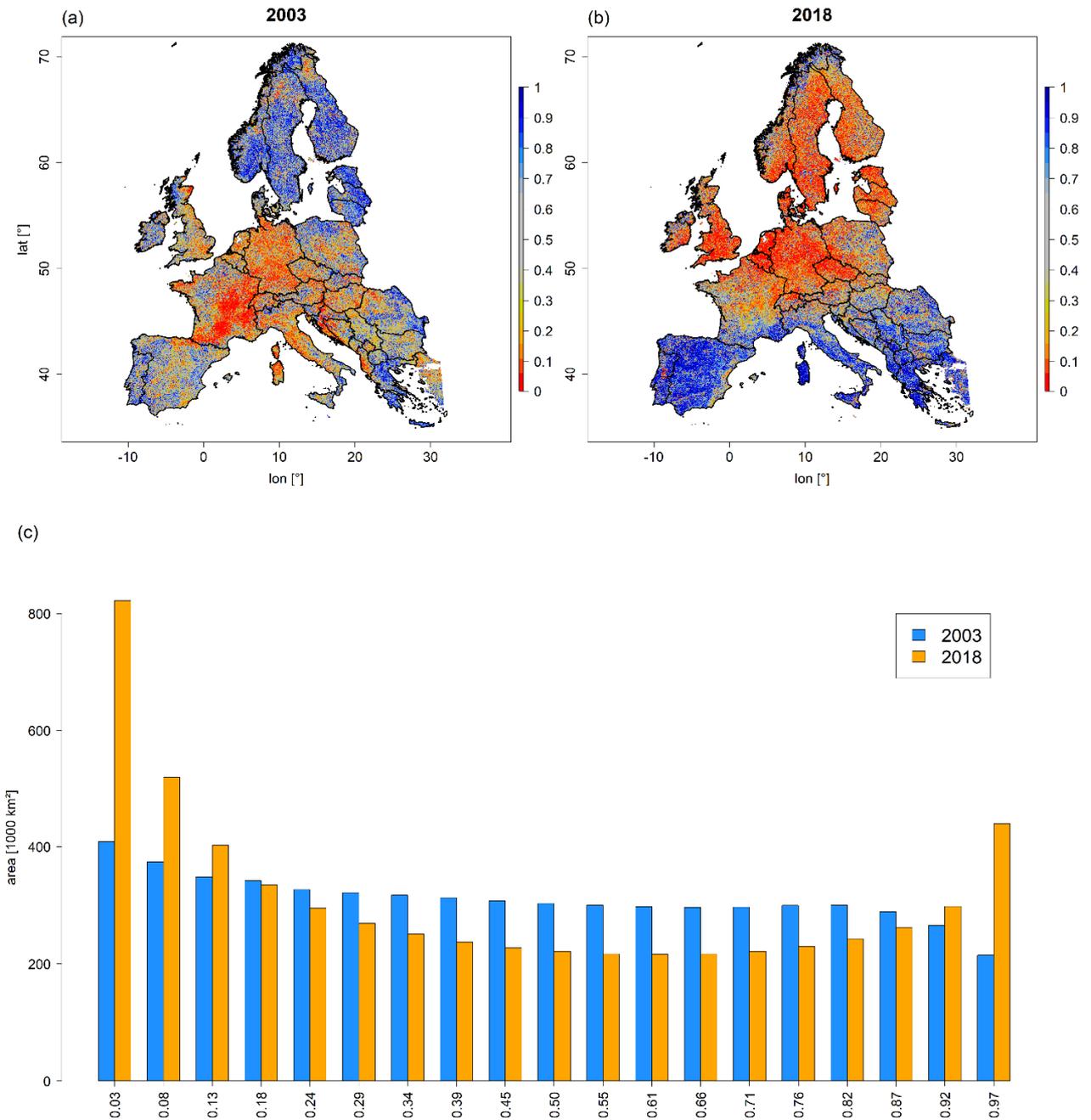

Fig. 4: MODIS NDVI quantiles representing peak-season conditions at the end of July (DOY 209) in 2003 (a) and 2018 (b) as well as the corresponding area histograms (in units of 1000 km²) representing the nineteen NDVI quantiles (c). Blue colors in (a) and (b) indicate upper quantiles (thus a higher than average vegetation greenness), while orange to red colors indicate lower anomalies (i.e. lower than average vegetation greenness). Blue bars in (c) refer to 2003 and orange bars to 2018. Similar results for MODIS EVI are shown in supplementary Fig. S5.

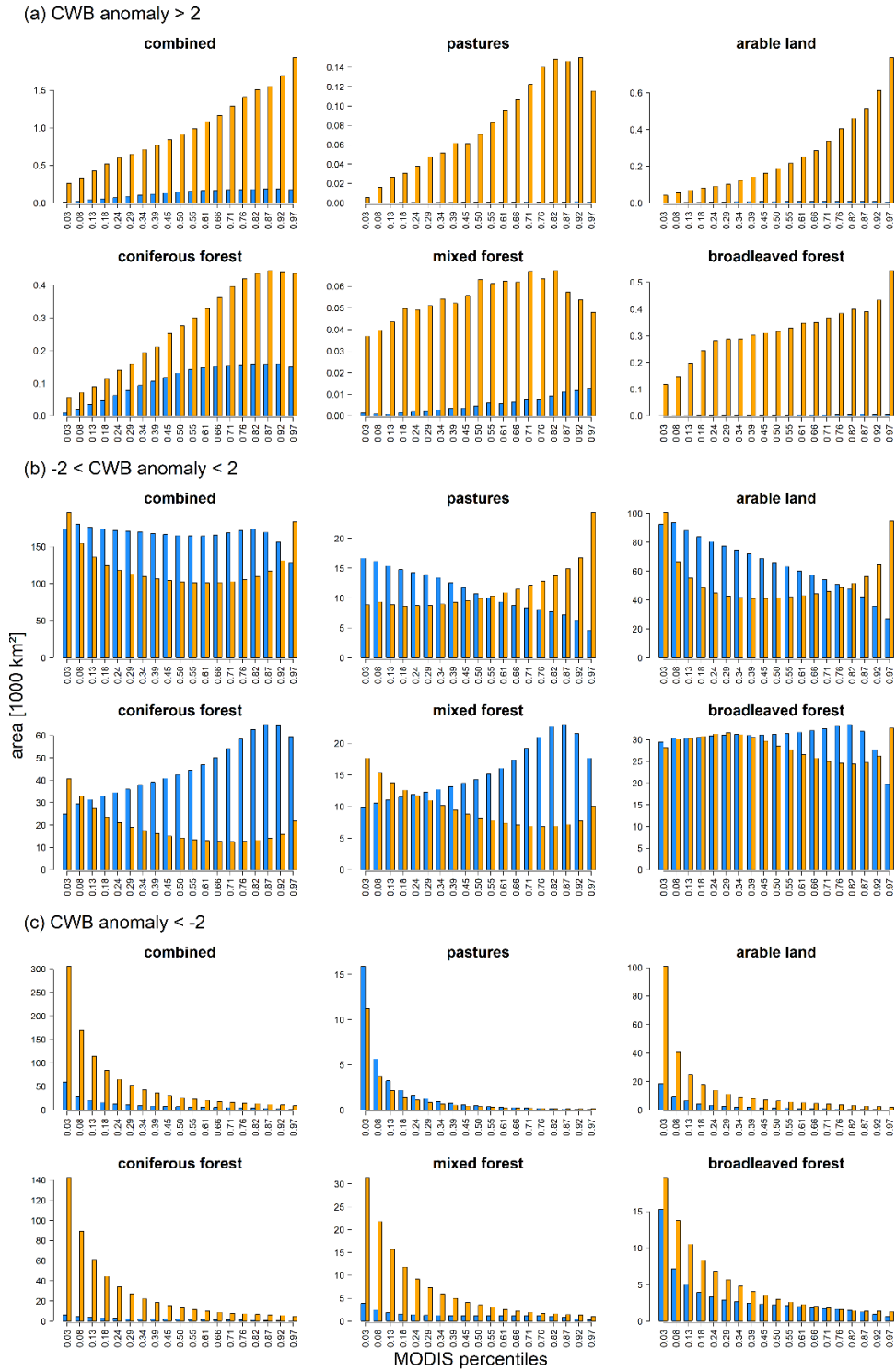

Fig. 5: Histograms depicting the absolute areas (in units of 1000 km²) representing the nineteen NDVI quantiles pooled according to CORINE land-cover classes for regions that featured (a) water surplus (CWB-anomaly > 2), (b) average conditions (- 2 < CWB-anomaly < 2), and (c) water deficit (CWB-anomaly < -2). Blue bars refer to 2003, orange bars to 2018. Similar results for MODIS EVI are shown in supplementary Fig. S6.

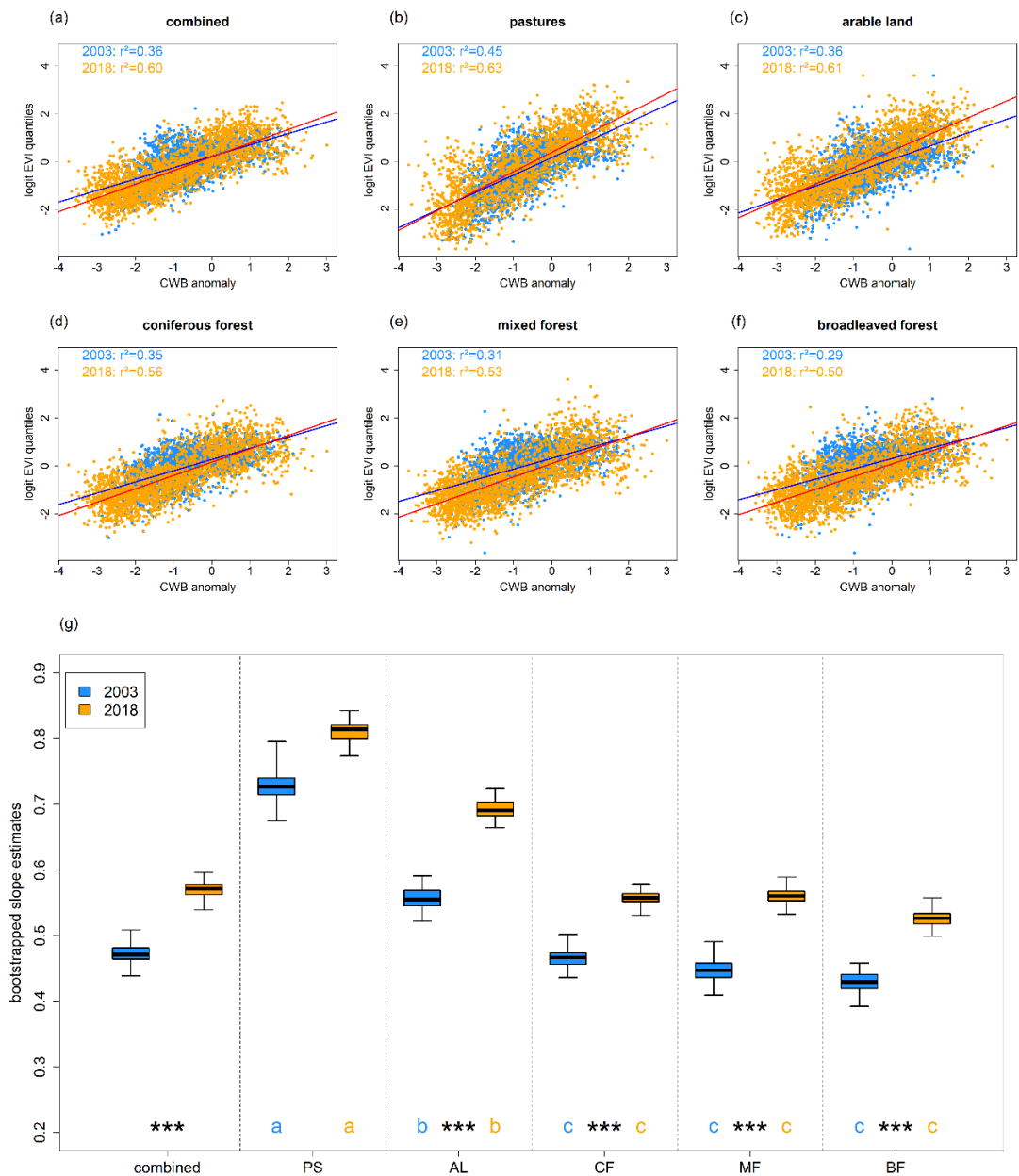

Fig. 6: (a-f) Scatterplots depicting the relationship between average logit-transformed NDVI-quantiles and mean CWB of the 100 CWB percentiles in 2003 (blue) and 2018 (orange) for pastures (b), arable land (c), coniferous forests (d), mixed forest (e), broadleaved forest (f) and a combination of those (a). Blue lines depict the regression line for 2003, red lines for 2018. (g) Bootstrapped regression slope estimates for the five different land-cover classes as well as their combination. Minor case letters refer to group assignment of land-cover classes according to the overlap of 99.9 % confidence intervals of bootstrapped slopes in 2003 (blue) and 2018 (orange). Significance stars (***) indicate no overlap between 99.9 % confidence intervals of 2003 and 2018 for the respective land-cover class. PS = pastures, AL = arable land, CF = coniferous forest, MF = mixed forest, BF = broadleaved forest. Similar results for MODIS EVI are shown in supplementary Fig. S8.

# 4 Discussion

## 4.1 Climatic framework

Based on the parameters considered for the quantification of summer conditions, 2018 clearly supersedes 2003 (Figs. 1-3). That is, the anomaly of integrated 500 hPa geopotential height indicated more persistent anticyclonic circulation patterns for 2018 across Northern Europe, which likely triggered the strong positive heat load anomalies and negative climatic water balances across Central and Northern Europe in comparison to 2003. Altogether, the area featuring strong positive heat load anomalies and strong negative climatic water balance anomalies was respectively 7.1 and 5.3 times larger in 2018, compared to 2003. However, another specific feature of 2018 in comparison to 2003 is the clear dipole of 500 hPa geopotential height, which resulted in lower heat loads and an above average climatic water balance in the Mediterranean.

Anticyclonic blocking situations – as indicated by strong positive 500 hPa geopotential height anomalies – have been reported an increasing frequency in course of the satellite era (Horton et al., 2015) which likely relates to the increasing persistence of heatwaves observed over the past 60 years (Pfleiderer and Coumou, 2018) and the increasing frequency of a hemisphere-wide wavenumber 7 circulation pattern (Kornhuber et al., 2019). The resulting heatwaves are additionally enhanced by global warming and positive land surface – atmosphere feedback loops via soil moisture depletion and subsequent lack of latent cooling (Fischer et al., 2007). Moreover, summer temperatures and precipitation were reported to correlate negatively at mid latitudes which may amplify further according to CMIP5 climate projections (Zscheischler and Seneviratne, 2017). This renders the evaluation of ecosystem responses to compound events a key topic for climate change research (Zscheischler et al., 2018). Consequently, the persistent climatic blocking situation which resulted in a large spatial distribution of extreme temperature and water balance anomalies qualifies 2018 as a key event for studying ecosystem-responses to anticipated hotter droughts in Europe.

## 4.2 Ecosystem impact

End of July MODIS NDVI anomalies generally supported the impression of a more extreme drought in 2018. That is, the area featuring the lowest quantile in 2018 was roughly twice as high if compared to 2003 (Fig. 4c). In accordance with the more northward location of the anticyclone, hotspots of ecosystem impact were concentrated in Central and Northern Europe in 2018. At the same time, large parts of the Mediterranean featured positive NDVI deviations in 2018 which resulted in a two times larger area represented by the highest NDVI quantile compared to 2003 (Fig. 4c).

In general, the spatial distribution of NDVI quantiles matches the observed climatic dipole of 2018 very well (see section 4.1). Moreover, the bimodal response of European ecosystems is well in line with a preliminary report on European maize yield in 2018, with an observed strong increase (10% more than average) for e.g. Romania and Hungary, a strong decrease (10% less than average) for e.g., Germany and Belgium, and the European level net effect estimated at a decrease in maize yield of around 6% (see public references: European Commission).

The bimodal behavior and larger extent of ecosystem impact in 2018 was also reflected in land-cover specific area distributions of NDVI quantiles (Fig. 5a and 5c). That is, the area featuring lowest (highest) quantiles in regions with water deficit (surplus) was much larger in 2018 compared to 2003. Given the skewed distribution of quantiles in those regions, i.e. more lower quantiles in regions with strong negative CWB anomalies and more upper quantiles in regions with strong positive CWB anomalies, the response of these ecosystems appears to be governed by prevailing climate conditions.

This impression was underlined by linear regressions, which revealed a strong and significant positive impact of CWB on NDVI quantiles in both years and for all land-cover types (Figs. 6a-f). It is noteworthy, that 2018 was characterized by a higher explained variance and significantly more positive model slopes in comparison to 2003 (Fig. 6g). These impressions from single linear models were generally supported by the linear mixed effects model, which showed an increase of $r^2$ at about 9 percent (from 0.47 to 0.51) when incorporating land-cover and year as random effects.

The stronger coupling between CWB and NDVI quantiles in 2018 may be related to the period used to integrate CWB (April-July). To test this, we assessed the relationship of NDVI quantiles with CWB integrated over various different periods (e.g. including previous winter in CWB or shortening the period) which all revealed similar patterns, i.e. a stronger response to CWB in 2018 (not shown). However, it seems possible that the stronger coupling of NDVI quantiles with CWB in 2018 is related to the spatial distribution of drought and thus the ecosystems being affected. In 2003, the epicenter of the drought was located in Central France and the Mediterranean, i.e. regions which host ecosystems that are regularly experiencing summer drought and thus likely are better adapted to dry conditions. In contrast, the circulation patterns of 2018 resulted in a drought-epicenter in Central Europe, Southern Scandinavia, and around the Baltic Sea, i.e. regions with ecosystems which are less adapted to extremely dry climatic conditions such as in 2018 and therefore likely react strongly. Various reports from dried-out pastures and cornfields as well as deciduous trees shedding their leaves in July and August likely explain the record-low NDVI values for corresponding Central European ecosystems (see public news references). At the same time, the usually summer-dry Mediterranean experienced water surplus, relaxing the general limitation of the associated ecosystems by plant water availability and leading to a generally higher greenness of the vegetation. Considering forest ecosystems, this interpretation is in in line with Klein (2014), who reported higher leaf gas exchange and thus photosynthetic capacity under dry conditions in Mediterranean forests compared to temperate forests. Nevertheless, our hypothesis explaining the stronger coupling of NDVI quantiles with CWB in 2018 needs further investigation, e.g. by studying the sensitivity and coupling of plant productivity with climatic properties for the considered land-cover classes as for instance done by Anderegg et al. (2018). A sub-classification of land-cover classes seems to be reasonable for such an analysis, in order to account for possibly differing drought-sensitivities of ecosystems represented by one specific land-cover class. For instance, Mediterranean coniferous forests are likely more adapted to drought conditions than boreal coniferous forests in Scandinavia.

In addition to the differences between the two drought events, we also observed differing sensitivities of NDVI quantiles to CWB among ecosystems, which were consistent over the two drought events. We found the highest regression slope estimates for pastures followed by arable land, coniferous forests, mixed forests, and broadleaved forests (Fig. 6g). This likely reflects the higher climatic buffering function of forests in comparison to arable land and pastures. In forests, the micro-climate is generally less extreme, leading to lower ambient air temperatures and consequently a lower evapotranspiration in comparison to open fields (Chen et al., 1993, 1999; Young and Mitchell, 1994). Consequently, water resources are consumed more sustainably by forests. Moreover, if not growing on water-logged soils trees typically feature higher rooting depths compared to grasses and crops and therefore have access to deeper soil water reservoirs. Regarding the European drought of 2003, an accelerated soil moisture depletion of grasslands in comparison to forests has been reported earlier (Teuling et al., 2010). Also Wolf et al. (2013) found contrasting responses of grasslands vs. forests regarding water and carbon fluxes during a drought event in 2011. They observed an immediate negative drought-impact on the productivity of managed grasslands while mixed and coniferous forests simply reduced transpiration and maintained GPP, thereby increasing their water-use efficiency and decreasing soil-water consumption (Wolf et al., 2013).

Nevertheless, European forests were in parts heavily affected by the drought 2018, as indicated by the depicted density distributions (Fig. 5c). For instance, 140,000 km² of drought affected coniferous forests featured the lowest quantile in 2018. First estimates for North Rhine-Westphalia (Northwest Germany) assume forest productivity losses of around 40% (see public references: Wald und Holz NRW). Moreover, 20,000 km² of deciduous trees featured the lowest quantile in 2018 which likely relates to the observed early leaf-shedding of deciduous trees in Central Europe (see public-news references). Besides direct impacts, carry-over effects are likely to be experienced in the following years: Evidence for delayed responses comes from remotely sensed vegetation activity in the aftermath of the 2003 event (Reichstein et al., 2007) and reports on observed canopy die-back of beech trees in spring 2019 (GfÖ-workshop on drought 2018 held on June 4th, 2019 in Basel as well as public news references). These effects could partly be due to legacy effects in tree response to drought (Anderegg et al., 2015) as well as tree mortality often occurring years after the event (Bigler et al., 2006; Cailleret et al., 2017). Support for an expected delayed response of forests also comes from a severe drought in Franconia, Southern Germany, in 2015, which resulted in increased Scots pine mortality, that was not recognized earlier than in the subsequent winter 2015/2016 and became even more pronounced in spring 2016 (Buras et al., 2018). From this event, we also learned that particularly forest edges – which feature an intermediate micro-climate between the forest interior and open fields – are more susceptible to drought-induced mortality (Buras et al., 2018). However, given the spatial resolution of the applied remote sensing products (231 m x 231 m) patches with tree dieback as well as forest edges could not be resolved.

Given likely legacy effects, studying the development of Central and Northern European forest ecosystems over the next years is particularly interesting and may reveal negative mid- to long-term responses as well as an increased tree mortality. In this context, we propose an immediate observation of forests within the outlined hotspots by combining satellite-based and close-range remote sensing techniques with dendroecological investigations and an eco-physiological monitoring (Buras et al., 2018; Ježík et al., 2015). A timely initiation of such monitoring campaigns would provide the unique opportunity to study natural tree die-back in real-time, thereby increasing our knowledge about drought-induced tree mortality (Cailleret et al., 2017).

## 4.3 A new reference for extreme drought in Europe?

Based on climatic evidence, 2018 may be considered the new reference year for hotter droughts in Europe. However, the observed contrasting patterns of the drought events in 2003 and 2018 highlight the complexity of ecosystem responses to severe droughts. More specifically, we observed a different sensitivity of ecosystems (as represented by VI quantiles) to CWB between the two events and that different land-cover classes revealed a differing sensitivity to drought, with pastures and agricultural fields expressing a higher sensitivity in comparison to forests. The observed climatic heterogeneity and resulting ecosystem response poses certain challenges for estimating the effects on the carbon cycle in European ecosystems in 2018. That is, the observed dipole in 2018 – with positive impacts in the Mediterranean and negative effects in Central and Northern Europe – makes it difficult to directly compare the European carbon budget of 2018 with 2003 (as done for 2003 in Ciais et al., 2005). Finally, legacy effects of forest ecosystems are likely to occur in course of the next years (Anderegg et al., 2015; Buras et al., 2018; Kannenberg et al., 2018) and already observed (see public news references). Consequently, to obtain a more complete picture about the impact of the drought 2018 we recommend continued satellite-based remote sensing surveys to be accompanied by immediate in-situ monitoring campaigns. In this context, particular attention should be given to the outlined hotspots of the drought 2018, i.e. Ireland, United Kingdom, France, Belgium, Luxemburg, the Netherlands, Northern Switzerland, Germany, Denmark, Sweden, Southern Norway, Czech Republic, Poland, Lithuania, Latvia, Estonia, and Finnland.

## 5 Conflict of Interest

The authors declare that the research was conducted in the absence of any commercial or financial relationships that could be construed as a potential conflict of interest.

## 6 Author Contributions

AB and CSZ developed the study design. AB conducted all data processing and statistical analyses. Interpretation and refinement of statistical results was discussed among AB, AR, and CSZ. AB drafted the first version of the article which was further refined by AR and CSZ.

## 7 Funding

This project is funded by the Bavarian Ministry of Science and the Arts in the context of the Bavarian Climate Research Network (BayKliF).

**Public news references**

https://www.climate.gov/news-features/event-tracker/hot-dry-summer-has-led-drought-europe-2018 (in English).

https://www.euronews.com/2018/08/10/explained-europe-s-devastating-drought-and-the-countries-worst-hit (in English).

First summary of 2018 drought-impacts on ecosystems in Switzerland: https://www.wsl.ch/de/newsseiten/2019/07/initiative-trockenheit-2018.html (in English).

Report on dying beech trees in the Hainich national park in spring 2019: https://www.nationalpark-hainich.de/de/aktuelles/aktuelles-presse/einzelansicht/extremjahr-2018-hinterlaesst-spuren-im-nationalpark-hainich.html

Short-term outlook of the European Commission for EU agricultural markets: https://ec.europa.eu/agriculture/markets-and-prices/short-term-outlook_en (in English).

MODIS-based maps on land surface temperature and cloud cover anomalies for whole Europe: https://www.geografiainfinita.com/2018/09/un-analisis-de-la-sequia-en-europa-en-el-verano-de-2018/?platform=hootsuite (in Spanish).

German atlas for soil water depletion, provided by the Umwelt-Forschungs-Zentrum UFZ: https://www.ufz.de/index.php?de=44429 (in German).

Report on the impacts of the heat-wave in Germany, provided by the Karlsruhe Institute of Technology KIT: http://www.kit.edu/kit/pi_2018_102_durre-betrifft-rund-90-prozent-der-flache-deutschlands.php (in German).

Interim report of ‚Wald und Holz NRW' on the impacts of the drought 2018 on forest productivity in North-Rhine Westphalia: https://www.wald-und-holz.nrw.de/aktuelle-meldungen/2018/zwischenbilanz-trockensommer-2018 (in German).

Pictures from dried-out cornfields in Germany: https://www.alamy.com/corn-field-dried-up-and-only-grown-low-small-corn-cobs-through-the-summer-drought-drought-in-ostwestfalen-lippe-germany-summer-2018-image215773502.html

Report on early leaf shedding of deciduous trees in Germany: https://www.wetteronline.de/wetternews/trockenheit-setzt-natur-zu-viele-baeume-werfen-ihr-laub-ab-2018-08-14-lb (in German)

**Data Availability Statement**

The datasets analyzed for this study are publicly available. Public web-links as well as the post-processing of data is described in the material and methods section of this article, wherefore all results are reproducible.

*Supplementary Material*

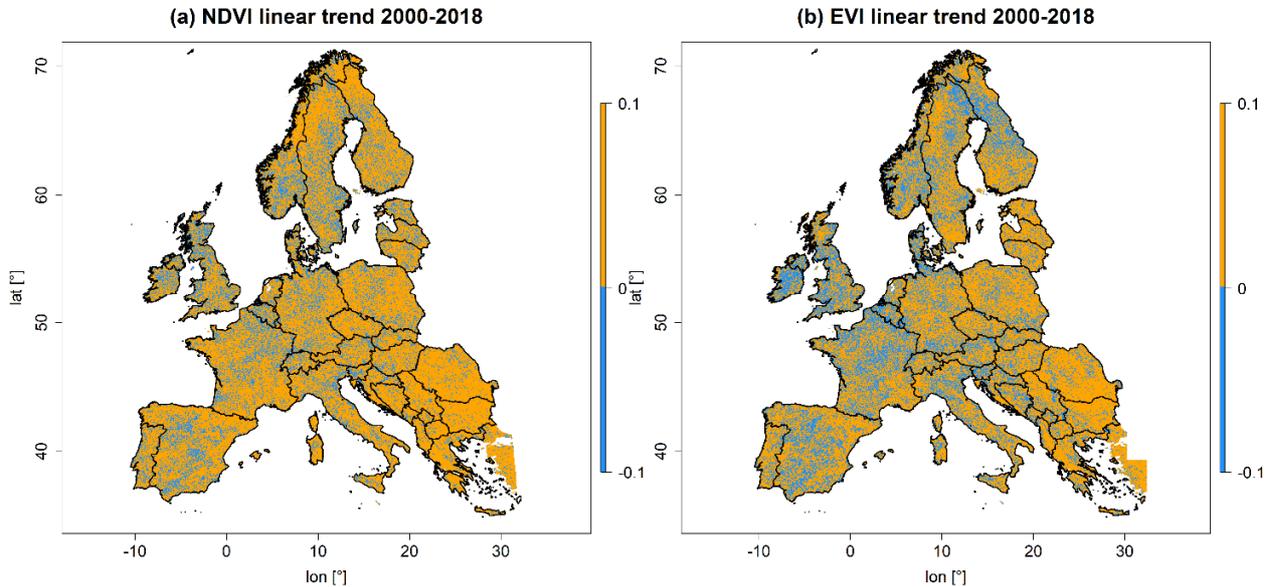

**Fig. S1:** Maps depicting temporal trends of NDVI (a) and EVI (b) over the whole study period. Individual trends were subtracted from each individual pixel time-series.

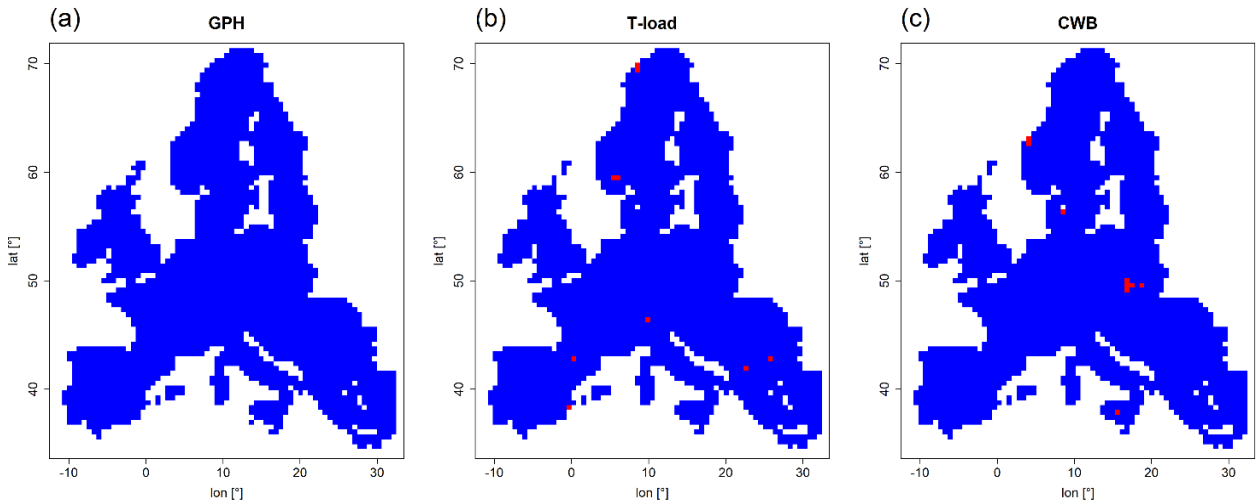

**Fig. S2:** Maps depicting significance of Shapiro-Wilk normality test (red pixels refer to $p < 0.001$) for 500 hPa geopotential height (a), heat load (b), and climatic water balance (c). The number of significant tests ranges from 0 percent for geopotential height to 0.3 percent for heat load and thus lies within the order of expected type I errors (0.1 percent).

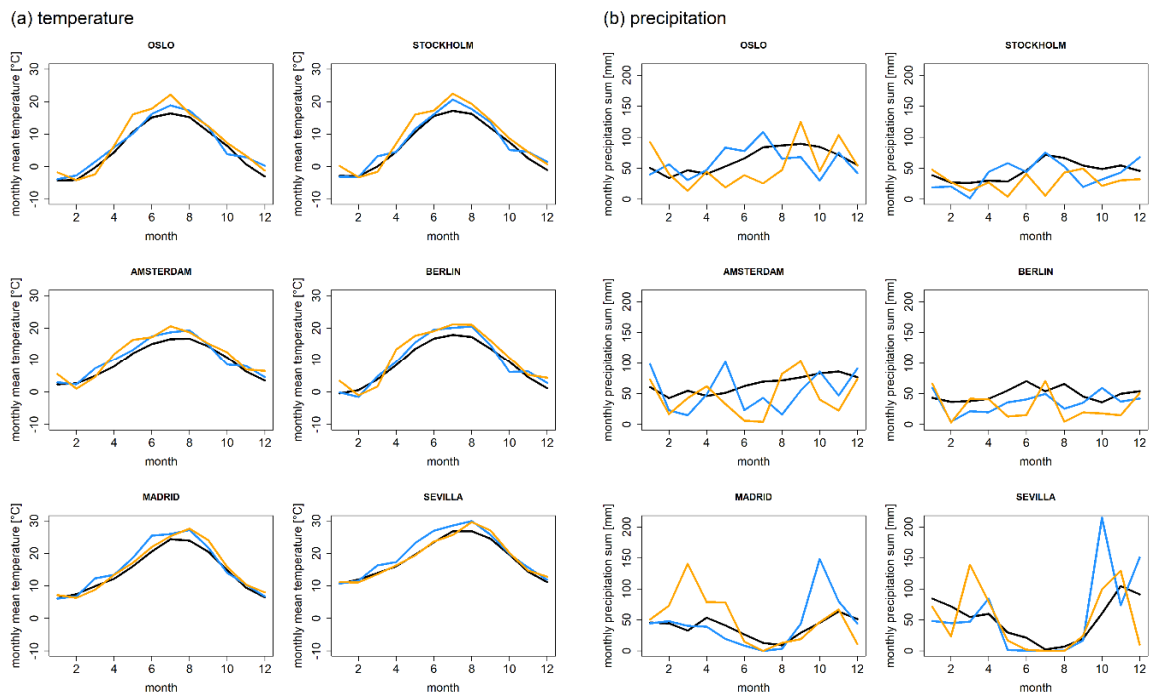

Fig. S3: Monthly temperature means (a) and precipitation sums (b) of six selected climate stations for 2003 (blue) and 2018 (orange) in comparison to mean values representative of the climate normal period 1961-1990.

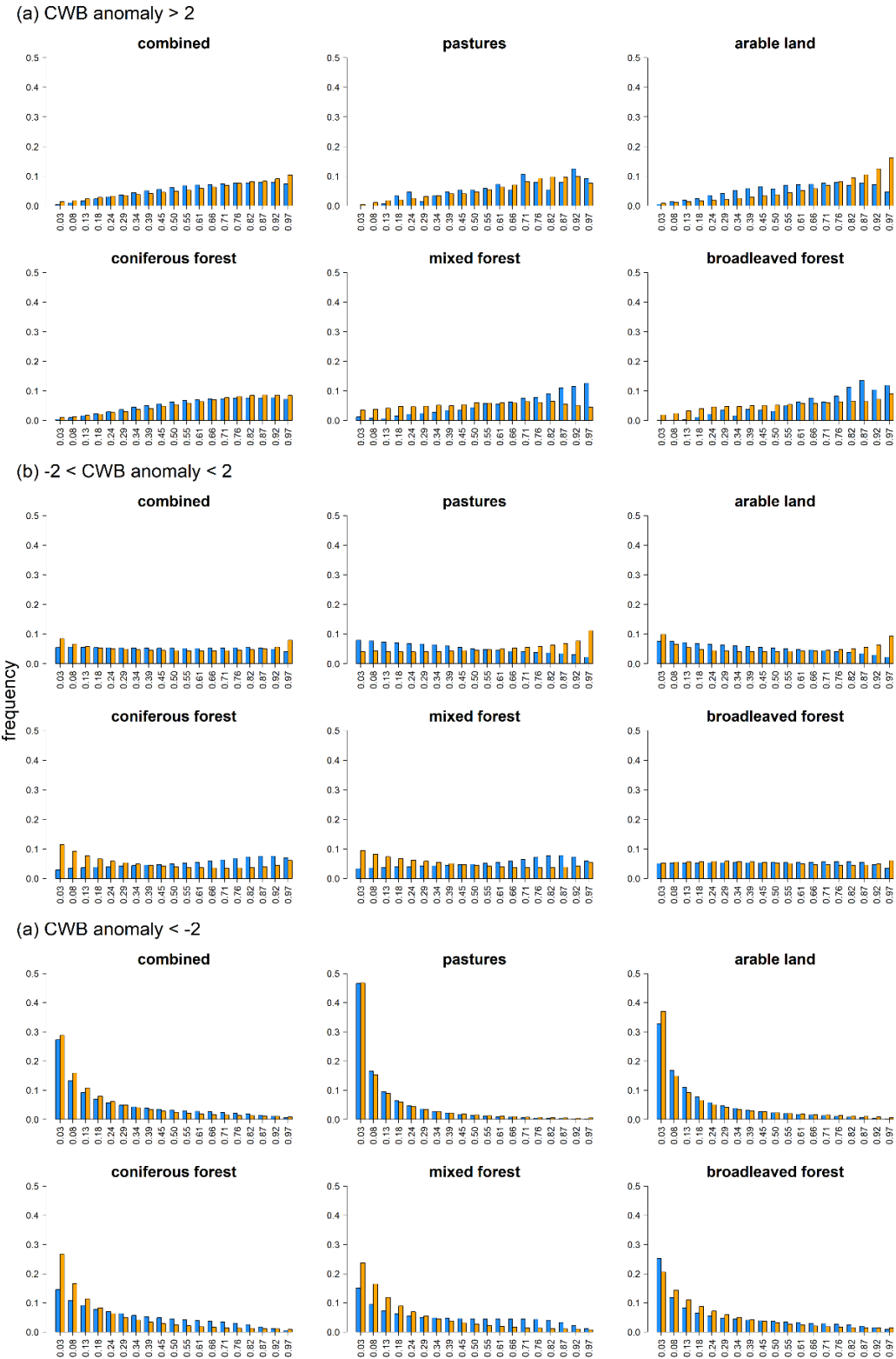

Fig. S4: Histograms depicting the proportions representing the nineteen NDVI quantiles pooled according to CORINE land-cover classes for regions that featured (a) water surplus (CWB-anomaly > 2), (b) average conditions (- 2 < CWB-anomaly < 2), and (c) water deficit (CWB-anomaly < -2). Blue bars refer to 2003, orange bars to 2018.

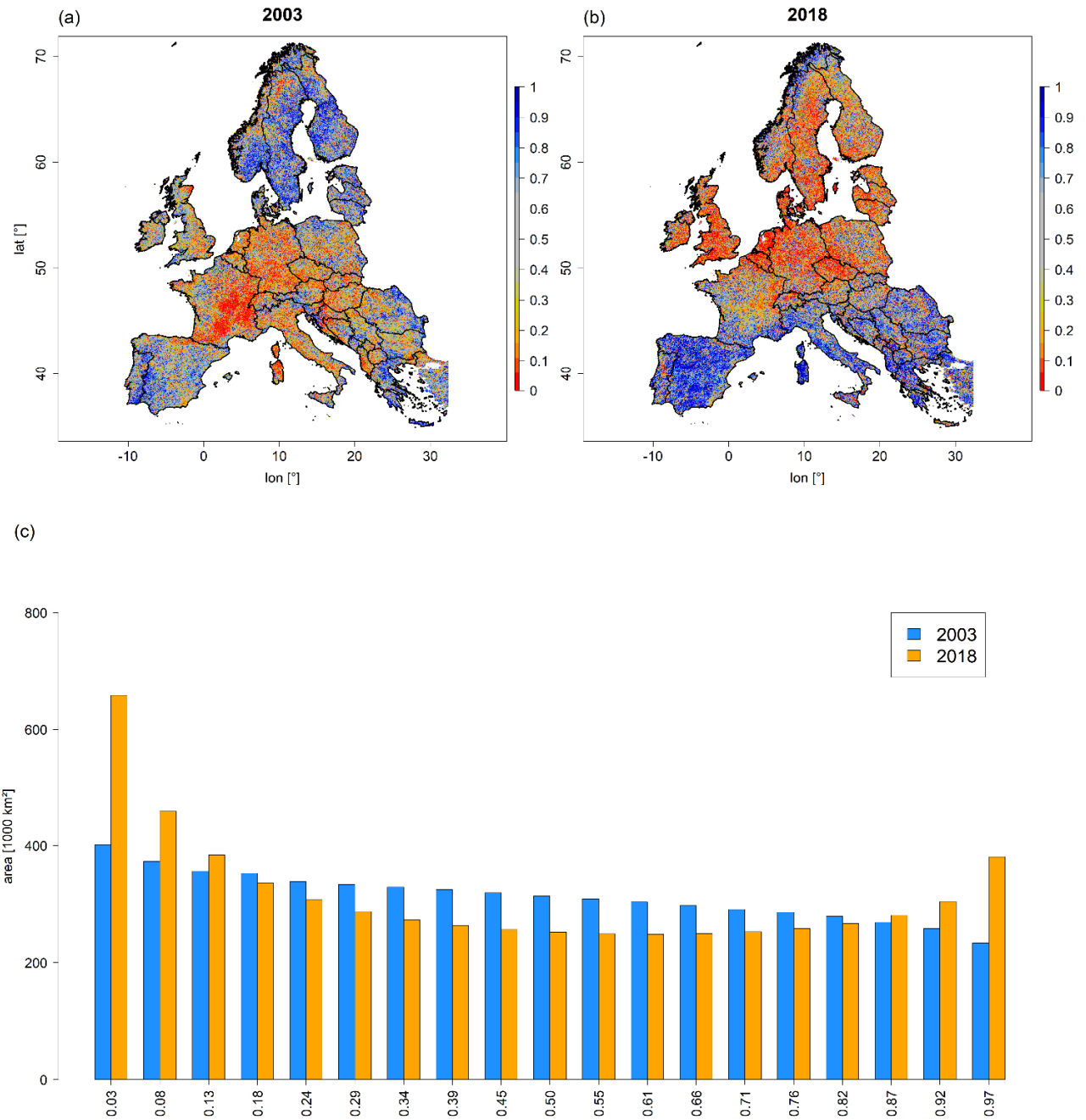

Fig. S5: MODIS EVI quantiles representing peak-season conditions at the end of July (DOY 209) in 2003 (a) and 2018 (b) as well as the corresponding area histograms (in units of 1000 km²) representing the nineteen EVI quantiles (c). Blue colors in (a) and (b) indicate upper quantiles (thus a higher than average vegetation greenness), while orange to red colors indicate lower anomalies (i.e. lower than average vegetation greenness). Blue bars in (c) refer to 2003 and orange bars to 2018.

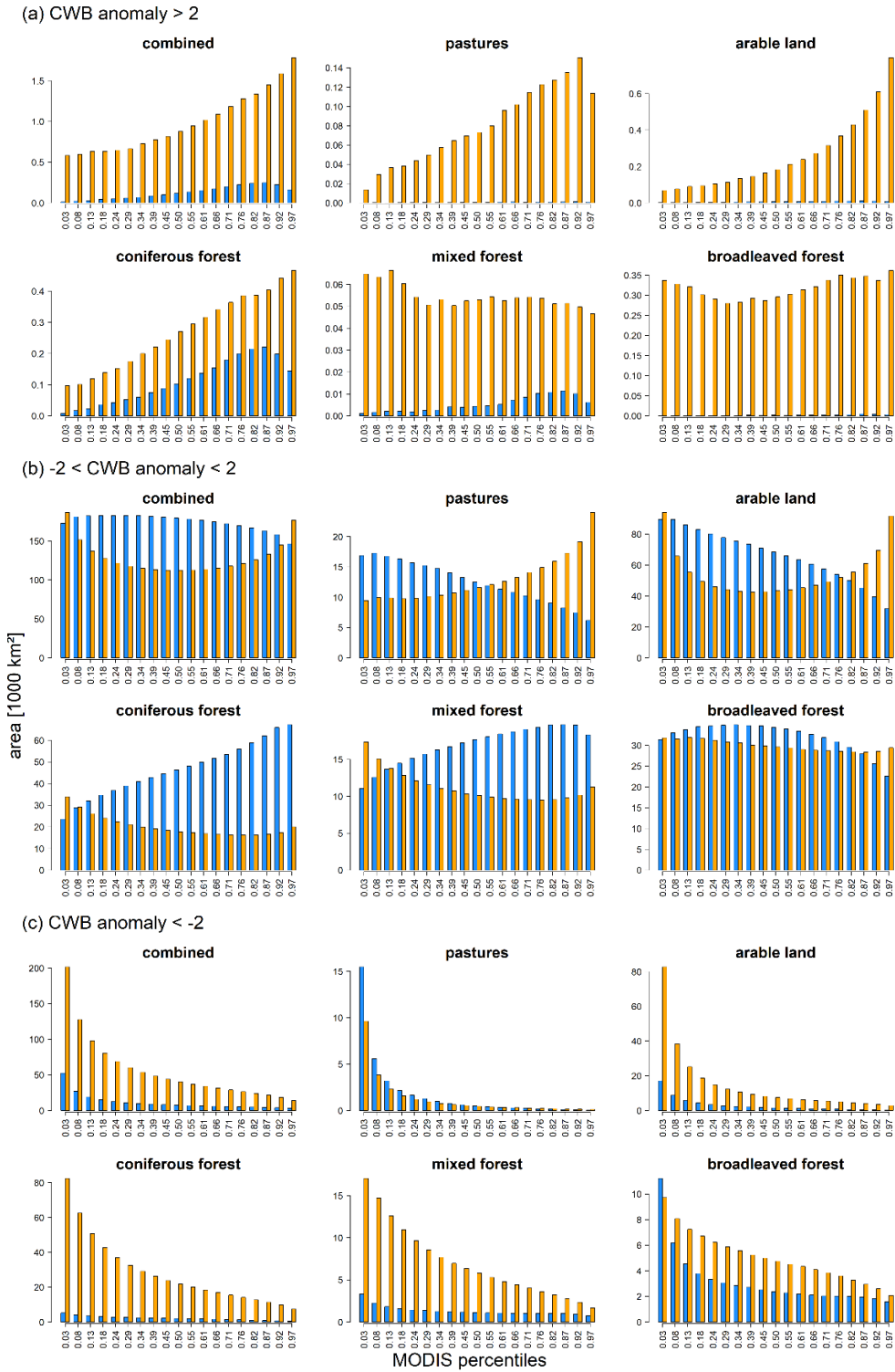

Fig. S6: Histograms depicting the absolute areas (in units of 1000 km²) representing the nineteen EVI quantiles pooled according to CORINE land-cover classes for regions that featured (a) water surplus (CWB-anomaly > 2), (b) average conditions (- 2 < CWB-anomaly < 2), and (c) water deficit (CWB-anomaly < -2). Blue bars refer to 2003, orange bars to 2018.

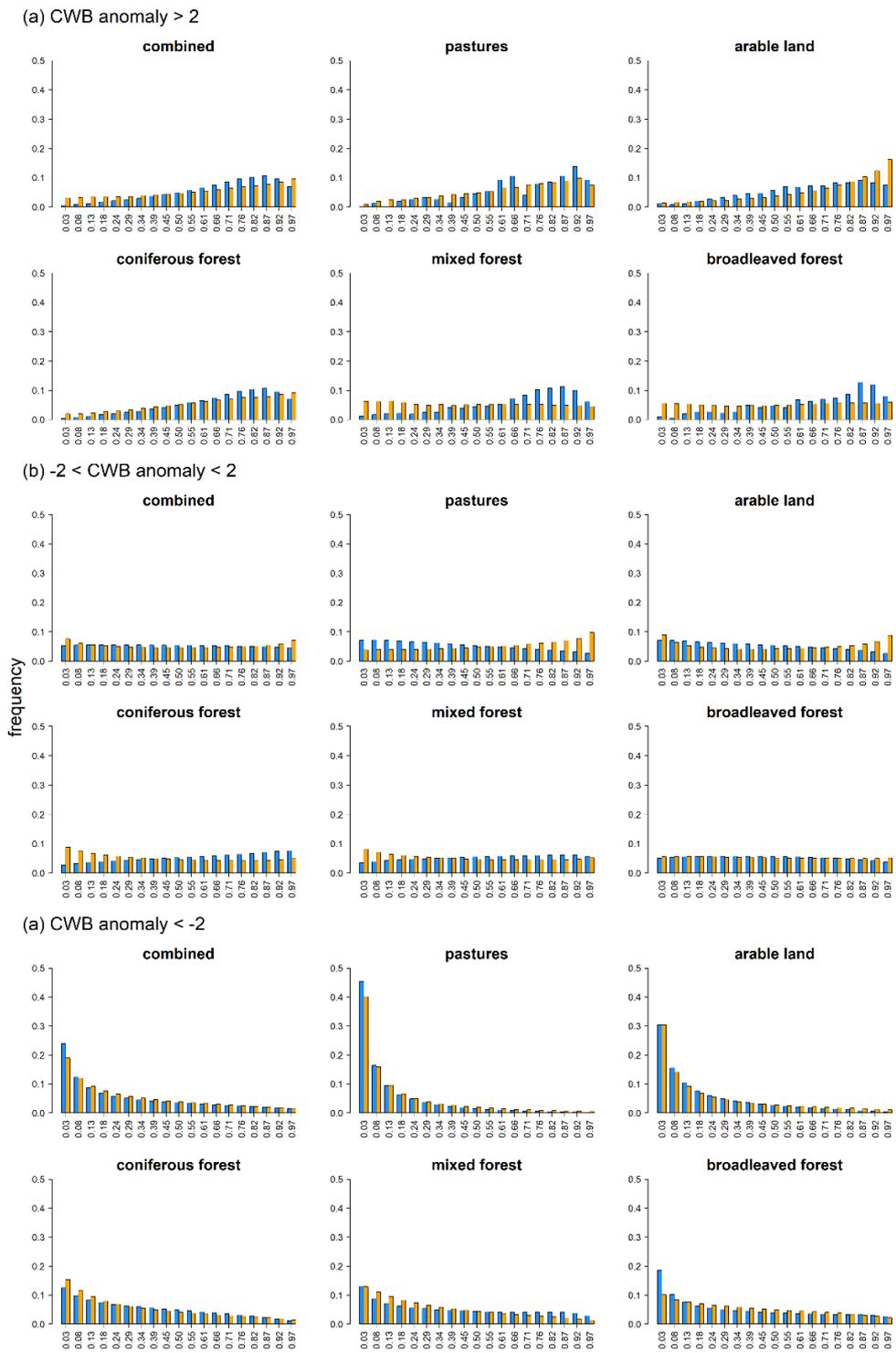

Fig. S7: Histograms depicting the proportions representing the nineteen EVI quantiles pooled according to CORINE land-cover classes for regions that featured (a) water surplus (CWB-anomaly > 2), (b) average conditions (- 2 < CWB-anomaly < 2), and (c) water deficit (CWB-anomaly < -2). Blue bars refer to 2003, orange bars to 2018.

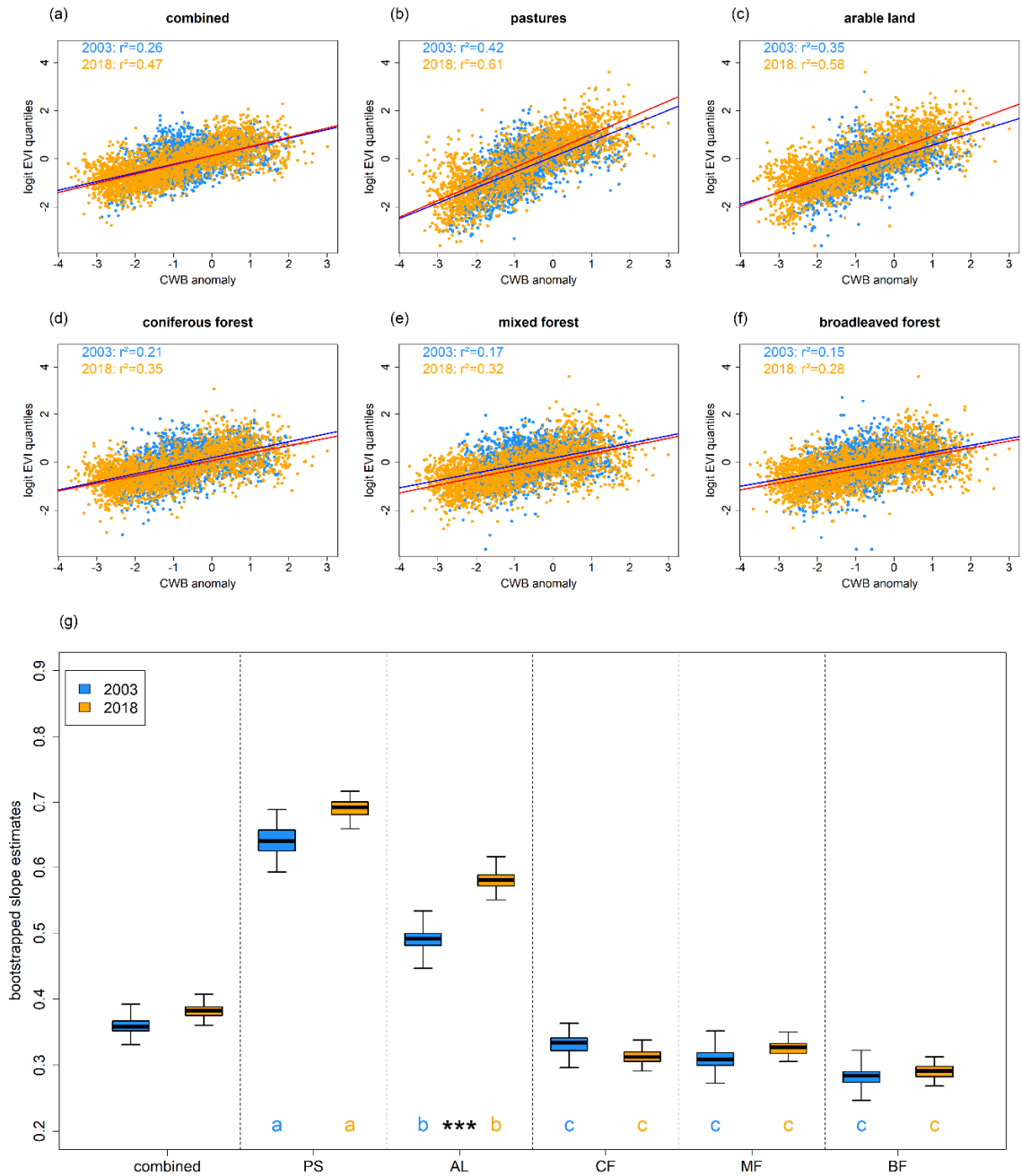

Fig. S8: (a-f) Scatterplots depicting the relationship between average logit-transformed EVI-quantiles and mean CWB of the 100 CWB percentiles in 2003 (blue) and 2018 (orange) for pastures (b), arable land (c), coniferous forests (d), mixed forest (e), broadleaved forest (f) and a combination of those (a). Blue lines depict the regression line for 2003, red lines for 2018. (g) Bootstrapped regression slope estimates for the five different land-cover classes as well as their combination. Minor case letters refer to group assignment of land-cover classes according to the overlap of 99.9 % confidence intervals of bootstrapped slopes in 2003 (blue) and 2018 (orange). Significance stars (***) indicate no overlap between 99.9 % confidence intervals of 2003 and 2018 for the respective land-cover class. PS = pastures, AL = arable land, CF = coniferous forest, MF = mixed forest, BF = broadleaved forest.